\documentclass[aps,prd,twocolumn,superscriptaddress,amsmath,amssymb,amsthm,nofootinbib, preprintnumbers]{revtex4}
\usepackage{url}
\usepackage[colorlinks=true,
    linkcolor=blue,
    citecolor=blue,
    urlcolor=blue]{hyperref}
 \usepackage{amsmath}
\usepackage{graphicx}
\usepackage{epstopdf}
\usepackage{float}
\usepackage{hyperref}
\usepackage{color}
\usepackage[T1]{fontenc}
\usepackage[utf8]{inputenc}
\usepackage[toc,page]{appendix}
\usepackage[usenames,dvipsnames]{xcolor}
\usepackage[normalem]{ulem}

\newcommand{\be}{\begin{equation}}
\newcommand{\ee}{\end{equation}}
\newcommand{\ba}{\begin{eqnarray}}
\newcommand{\ea}{\end{eqnarray}}

\newcommand{\beq}{\begin{equation}}
\newcommand{\eeq}{\end{equation}}
\newcommand{\beqa}{\begin{eqnarray}}
\newcommand{\eeqa}{\end{eqnarray}}

\begin{document}


\title{Electromagnetized black holes and swirling backgrounds in nonlinear electrodynamics: The ModMax case}

\author{Jos{\'e} Barrientos}
\email{jbarrientos@academicos.uta.cl}
\affiliation{Sede Esmeralda, Universidad de Tarapac{\'a}, Avenida Luis Emilio Recabarren 2477, Iquique, Chile}
\affiliation{Institute of Mathematics of the Czech Academy of Sciences, {\v Z}itn{\'a} 25, 115 67 Praha 1, Czech Republic}

\author{Adolfo Cisterna}
\email{adolfo.cisterna@mff.cuni.cz}
\affiliation{Sede Esmeralda, Universidad de Tarapac{\'a}, Avenida Luis Emilio Recabarren 2477, Iquique, Chile}
\affiliation{Institute of Theoretical Physics, Faculty of Mathematics and Physics,
Charles University, V Hole{\v s}ovi{\v c}k{\' a}ch 2, 180 00 Praha 8, Czech Republic}

\author{Mokhtar Hassaine}
\email{hassaine@inst-mat.utalca.cl}
\affiliation{Instituto de Matem\'{a}ticas,
Universidad de Talca, Casilla 747, Talca, Chile}

\author{Konstantinos Pallikaris}
\email{konstantinos.pallikaris@ut.ee}
\affiliation{Laboratory of Theoretical Physics, Institute of Physics, University of Tartu, W. Ostwaldi 1, 50411 Tartu, Estonia}



\begin{abstract}

This work focuses on constructing electromagnetized black holes and vortex-like backgrounds within the framework of the ModMax theory--the unique nonlinear extension of Maxwell's theory that preserves conformal symmetry and electromagnetic duality invariance. We begin by constructing the Melvin-Bonnor electromagnetic universe in ModMax through a limiting procedure that connects the spacetime of two charged accelerating black holes with that of a gravitating homogeneous electromagnetic field. 
Building on this result, we proceed to construct the Schwarzschild and C-metric Melvin-Bonnor black holes within the ModMax theory, representing the first black hole solutions embedded in an electromagnetic universe in the context of nonlinear electrodynamics. While the characteristics of the Melvin-Bonnor spacetime and some of its black hole extensions have been widely examined, we demonstrate for the first time that the Schwarzschild-Melvin-Bonnor configuration exhibits an unusual Kerr-Schild representation. Following this direction, we also unveil a novel Kerr-Schild construction for the spacetime of two accelerating black holes, drawing on the intrinsic relationship between the Melvin-Bonnor spacetime and the C-metric. Finally, we expand the spectrum of exact gravitational solutions within Einstein-ModMax theory by constructing a vortex-like background that coexists with the Melvin-Bonnor universe. In this process, the Taub-NUT spacetime in ModMax has played a crucial role. We present this Taub-NUT solution in a different gauge that facilitates the comparison with the Melvin-Bonnor-Swirling case.

\end{abstract}

\maketitle

\section{Introduction}

Classical modifications to Maxwell's theory that become significant in strong field regimes, while naturally reducing to Maxwell's electrodynamics in the weak field approximation, are known as nonlinear theories of electrodynamics (NLE) \cite{Sorokin:2021tge}. 
Historically, they have been inspired by the development of Born-Infeld \cite{Born:1934gh} and Euler-Heisenberg \cite{Heisenberg:1936nmg} theories. Born-Infeld theory aimed to address the infinite self-energy in the electron's electric field, whereas Euler-Heisenberg theory provided a complete nonperturbative one-loop effective action for Quantum Electrodynamics, accounting for vacuum polarization effects due to virtual electrons and positrons. 
Being the electromagnetic duality of Maxwell field equations and the conformal invariance of Maxwell action some of the most remarkable features of Maxwell theory in $4d$, NLE have ultimately evolved to produce a nonlinear generalization of Maxwell's theory that preserves both of these symmetries. Dubbed ModMax theory \cite{Bandos:2020jsw, Kosyakov:2020wxv, Sorokin:2021tge}, the recently proposed model represents a one-parameter extension of Maxwell's theory in four dimensions. 
Several aspects of this model have come under scrutiny in the last years, see e.g. \cite{Bandos:2021rqy, Neves:2022jqq, Guzman-Herrera:2023zsv, Pantig:2022gih, Lechner:2022qhb, Ferko:2023iha, Ferko:2022cix,  Ferko:2023ozb, Rathi:2023vhw, Colipi-Marchant:2023awk, Garcia:2022wad, Deger:2024jfh}. Naturally, investigating the black hole spectrum of the theory, in essence, exact solutions to the Einstein-ModMax field equations, serves as a guide to better understand the effects that arise from nonlinearities of the ModMax theory. 

In electrovacuum, black hole spacetimes are characterized by seven parameters being the mass $m$, the Kerr-like rotation parameter $a$, the electric and magnetic charges $q_e$ and $q_m$, the cosmological constant $\Lambda$, and the more exotic NUT and 
acceleration parameters $n$ and $A$, respectively. The largest spacetime involving all these parameters is the Pleban\'ski--Demia\'nski configuration \cite{Plebanski:1976gy, Podolsky:2021zwr, Podolsky:2022xxd} which is the most general algebraically special solution of type D of the Einstein-Maxwell equations, and its recent generalization of algebraically general nature (type I), dubbed as Enhanced Pleban\'ski--Demia\'nski \cite{Barrientos:2023dlf}.   Additionally, two specific scenarios that fall outside these constructions involve the consideration of external electromagnetic and vortex-like fields. These scenarios correspond to Melvin-Bonnor geometries \cite{Bonnor_1954, Melvin:1963qx} and swirling geometries \cite{Gibbons:2013yq, Astorino:2022aam,Capobianco:2023kse}, both of which can accommodate the embedding of the families of solutions mentioned above. Their construction generally requires the use of a certain group of Lie point symmetries of the electrovacuum, specifically, Harrison and Ehlers symmetries of the magnetic type \cite{harrison1968new, Ehlers:1959aug}. 

Up to now, the study of black hole spacetimes within ModMax theory has primarily focused on spherically symmetric configurations, see e.g.  \cite{Flores-Alfonso:2020euz, Amirabi:2020mzv}. However, there are two notable exceptions: a stationary solution characterized by Taub-NUT geometry \cite{BallonBordo:2020jtw, Flores-Alfonso:2020nnd} and an accelerating black hole solution \cite{Barrientos:2022bzm}, with the latter being the first C-metric spacetime identified in a NLE model. Conversely, rotating solutions appear to fully engage the nonlinear aspects of the theory, as no such geometries have been constructed to date, even within the slowly rotating approximation \cite{Kubiznak:2022vft}.

The main objective of this work is to expand the spectrum of black hole solutions within the ModMax theory and to gain deeper insight into the technical challenges that prevent the construction of spacetimes with certain characteristics. 
Considering the current catalog of solutions in Einstein-ModMax theory and the persistent challenge of constructing rotating configurations, we focus on the remaining set of spacetimes characterized by background electromagnetic and vortex-like fields, specifically the Melvin-Bonnor and swirling geometries. Although Melvin-Bonnor spacetimes have been previously explored within the context of NLE, these constructions have faced limitations, particularly when a nontrivial cosmological constant is involved \cite{Gibbons:2001sx}. Moreover, no black hole immersed in a Melvin-Bonnor spacetime has yet been realized in any NLE theory whatsoever. On the other hand, swirling geometries have not been investigated in the context of NLE, even at the background level without an immersed black hole. The study of such a geometry is especially intriguing, as even if the swirling rotation is not of the Kerr-type, massive bodies do not drag the spacetime via its rotation, but an intrinsic stationarity of the spacetime itself drags massive bodies instead, these geometries may provide valuable insights for the future development of standard rotating spacetimes within the theory.

To construct Melvin-Bonnor geometries, we employ the innovative procedure proposed in \cite{Havrdova:2006gi, Havrdovathesis}. This latter represents a self-gravitating extension of the well-known concept that a homogeneous electric field in Minkowski spacetime is easily retrieved by taking Born's solutions describing two charged accelerating test particles and enlarging the distance between them while properly increasing the value of the charges. 
Therefore, following this approach, we first construct the Melvin-Bonnor-ModMax universe by starting from the spacetime of two accelerated charged black holes, specifically the C-metric spacetime discovered in \cite{Barrientos:2022bzm}. By examining the near-horizon geometry around the Rindler horizon and sending both black holes to infinity, we obtain the homogeneous, self-gravitating electromagnetic background characteristic of the Melvin-Bonnor-ModMax spacetime. Notably, unlike the analogous procedure in flat spacetime, the charge of the resulting configuration remains finite. Furthermore, this method naturally incorporates a nontrivial cosmological constant without requiring fine-tuning of the external electromagnetic field. Building on this foundation, an informed conjecture allows us to construct the Schwarzschild-Melvin and C-metric-Melvin black holes within ModMax theory, a result that could potentially indicate the existence of an underlying electromagnetization symmetry in the Einstein-ModMax framework. 
In the specific case of these configurations, such symmetry can be attributed to the electromagnetization symmetries inherent to the Einstein-Maxwell model. As we will show for the Melvin-Bonnor, Schwarzschild-Melvin, and C-metric-Melvin spacetimes, the electromagnetic invariants that define the ModMax action simplify, allowing the action to ultimately reduce to that of Maxwell's theory.

Furthermore, we begin the exploration of charged spacetimes with a background vortex-like field, focusing on a swirling background geometry for the ModMax theory. While this configuration does not exhibit a Kerr-type rotation, it represents the first example of a highly nontrivial stationary spacetime in ModMax theory. A key aspect of its construction is the identification of the associated magnetic field. Similar to the Taub-NUT case \cite{BallonBordo:2020jtw, Flores-Alfonso:2020nnd}, the nonlinearity of the theory becomes pronounced when the magnetic field includes a monopole contribution. Consequently, the field equations can be directly integrated by considering the magnetic field generated by the interaction between the spacetime’s stationarity and a monopole electric charge. 
In our pursuit of constructing a vortex-like background solution, we first revisit the Taub-NUT-ModMax spacetimes \cite{BallonBordo:2020jtw, Flores-Alfonso:2020nnd}, writing the solutions in an \textit{ad hoc} gauge useful for our purposes.  
For the swirling spacetime, we utilize an external electromagnetic field of the Melvin-Bonnor type. The stationarity of this spacetime allows the magnetic field component to generate an external electric field, thereby simplifying the integration of the field equations in analogy with the Taub-NUT scenario. 
A swirling Melvin-Bonnor background spacetime is thus obtained. While an exact solution for the black hole case has not yet been found, we suggest that the main challenge is computational in nature. This hints at the potential existence of another underlying Lie point symmetry, this time of the Ehlers type, in Einstein-ModMax theory. This symmetry, unlike the electromagnetic one, should be entirely nonlinear. The stationarity of the spacetime prevents the ModMax action from resembling Maxwell's theory, as there is no particular simplification of the electromagnetic invariants in these configurations.

The plan of the paper is divided as follows. In Section \ref{secII}, we provide a concise introduction to ModMax theory and its field equations, together with a review of the accelerating ModMax black hole solutions discovered in \cite{Barrientos:2022bzm}. In Section \ref{secIII}, we describe and apply the limiting procedure used to derive the Melvin-ModMax spacetime from the ModMax-C-metric, with a brief discussion of the key properties of this electromagnetic background. We also present a generalization that incorporates a nontrivial cosmological constant. In Section \ref{secIV}, we extend this background solution to include cases where a static or accelerating black hole is immersed in the Melvin-ModMax background. 
While these configurations in Einstein-Maxwell theory have been extensively studied, we present two new results: a novel Kerr-Schild representation for the Schwarzschild-Melvin-Bonnor spacetime and the Kerr-Schild representation for the spacetime of two charged accelerating black holes, specifically the charged C-metric. These two findings pertain to the ModMax case, but they are equally applicable to the Einstein-Maxwell framework in a straightforward manner.
In Section \ref{secV}, we focus on constructing a swirling background geometry. We begin by highlighting its similarities to the Taub-NUT case and 
re-writing explicitly the magnetic monopole contribution of these Taub-NUT spacetimes. 
This sets the stage for the presentation of a novel electromagnetic Melvin-Bonnor-Swirling background configuration. 
Finally, in Section \ref{secVI}, we conclude by summarizing our findings and outlining potential directions for future research on these geometries.
\section{ModMax theory and accelerating black holes}\label{secII}
We consider the Einstein-ModMax action principle given by
\begin{equation}
    \mathcal{I}=\frac{1}{16\pi}\int d^4x\sqrt{-g}(R-4\mathcal{L}),
\end{equation}
where $R$ is the Ricci scalar  and  $\mathcal{L}$ stands for the ModMax Lagrangian density whose expression reads
\begin{equation}
    \mathcal{L}=\frac{1}{2}(\mathcal{S}\cosh\gamma-\sqrt{\mathcal{S}^2+\mathcal{P}^2}\sinh\gamma).
\end{equation}
Here, $\mathcal{S}$ and $\mathcal{P}$ represent the simplest scalar and pseudo scalar electromagnetic invariants that can be constructed from the Maxwell-Faraday tensor $\mathcal{F}_{\mu\nu}=\partial_\mu \mathcal{A}_\nu-\partial_\nu \mathcal{A}_\mu$ and its dual $*\mathcal{F}_{\mu\nu}=\frac{1}{2}\epsilon_{\mu\nu\lambda\rho}\mathcal{F}^{\lambda\rho}$, namely, 
\begin{equation}
    \mathcal{S}=\frac{1}{2}\mathcal{F}_{\mu\nu}\mathcal{F}^{\mu\nu}, \quad \mathcal{P}=\frac{1}{2}\mathcal{F}_{\mu\nu}^{}*\mathcal{F}^{\mu\nu}.
\end{equation}
The Lagrangian density depends on $\mathcal{P}^2$; therefore parity invariance is not compromised. The parameter $\gamma$ is a dimensionless coupling constant restricted to $\gamma\geq0$; thus, well-posedness is ensured being causality and unitarity achieved \cite{Bandos:2020jsw}. In addition, since $\mathcal{L}$ is a convex function of the electric field, its energy-momentum tensor respects weak, strong, and dominant energy conditions \cite{Sorokin:2021tge}. 

Variations with respect to the metric and gauge fields yield the Einstein-ModMax field equations 
\begin{equation}
    G_{\mu\nu}=8\pi T_{\mu\nu},\quad d*\mathbf{E}=0,\quad d\mathcal{F}=0,
\end{equation}
where $\mathbf{E}=\mathbf{E}(\mathcal{F},*\mathcal{F})$ represents a nonlinear function of the Maxwell-Faraday tensor and its dual 
\begin{equation}
    \mathbf{E}_{\mu\nu}=\frac{\partial\mathcal{L}}{\partial \mathcal{F}^{\mu\nu}}=2(\mathcal{L}_\mathcal{S}\mathcal{F}_{\mu\nu}+\mathcal{L}_\mathcal{P}*\mathcal{F}_{\mu\nu}),
\end{equation}
and $T_{\mu\nu}$ the following the energy-momentum tensor 
\begin{equation}
    8\pi T_{\mu\nu}=4 \mathcal{F}_{\mu\sigma}\mathcal{F}_{\nu}^{\sigma}\mathcal{L}_\mathcal{S}+2(\mathcal{P}\mathcal{L}_\mathcal{P}-\mathcal{L})g_{\mu\nu}. \label{tmunumodmax}
\end{equation}
We make use of the shorthand notation $\mathcal{L}_{\mathcal{S}}=\partial\mathcal{L}/\partial\mathcal{S}$ and $\mathcal{L}_{\mathcal{P}}=\partial\mathcal{L}/\partial\mathcal{P}$. The function $\mathbf{E}(\mathcal{F},*\mathcal{F})$ is manifestly nonanalytic, as can be observed from the explicit expression  
\begin{equation}
    \mathbf{E}_{\mu\nu}=\left(\cosh\gamma-\frac{\mathcal{S}\sinh\gamma}{\sqrt{\mathcal{S}^2+\mathcal{P}^2}}\right)\mathcal{F}_{\mu\nu}-\frac{\mathcal{P}\sinh\gamma}{\sqrt{\mathcal{S}^2+\mathcal{P}^2}}*\mathcal{F}_{\mu\nu}, \label{EqModMax}
\end{equation}
and hence is ill-defined for null electromagnetic configurations. Notwithstanding, the Hamiltonian formulation of ModMax does support null configurations, showing at the same time how flat spacetime naturally belongs to the spectrum solution of the theory. Due to the conformal invariance $g\rightarrow\Omega^2 g$, being $\Omega$ an arbitrary function of the spacetime coordinates, Maxwell's theory does not form part of the weak field limit of ModMax theory; however, it is recovered in the $\gamma\rightarrow0$ limit \cite{Sorokin:2021tge}. This, up to some extend, is implied by the nonexistence of null field solutions in the Lagrangian formulation of the theory. Finally, the field equations are easily proven to be invariant under the electromagnetic duality transformation  
\begin{equation}
 \begin{pmatrix}
\mathbf{E}_{\mu\nu}^{\prime} \\
*\mathcal{F}_{\mu\nu}^{\prime} 
\end{pmatrix} = 
\begin{pmatrix}
\cos\theta & \sin\theta \\
-\sin\theta & \cos\theta 
\end{pmatrix}
 \begin{pmatrix}
\mathbf{E}_{\mu\nu}\\
*\mathcal{F}_{\mu\nu} 
\end{pmatrix}, 
\end{equation}
henceforth showing that the  ModMax theory is the unique NLE model sharing both conformal and electromagnetic duality invariance. 

It is direct to recognize that purely electric or magnetic solutions of Maxwell theory, namely, those for which $\mathcal{P}=0$, are going to be solutions of ModMax as well. However, the situation is different for $\mathcal{P}\neq0$. Already, for simple static and spherically symmetric spacetimes a deviation, although small, is retrieved. This is the case for Reissner-Nordstr\"om black holes in ModMax \cite{Flores-Alfonso:2020euz, Amirabi:2020mzv}. Pertinent to the first part of this work is the C-metric spacetime constructed in \cite{Barrientos:2022bzm}. Having an evident Maxwell-like gauge field profile, its main features and causal structure coincide with those of the standard geometry of a pair of charged accelerating black holes in electrovacuum. In spherical-like coordinates, and noticing the introduction of a cosmological constant $\Lambda=-\frac{3}{\ell^2}$, the solution is given by the following spacetime line element
\begin{equation}\label{Cmetricmodmax}
ds^2=\frac{1}{\Omega^2}\Bigl(-fdt^2+\frac{dr^2}{f}+r^2\Bigr[\frac{d\theta^2}{h}+h\sin^2\!\theta \frac{d\varphi^2}{K^2}\Bigr]\Bigr)\,, 
\end{equation}
where 
\begin{equation}
\begin{aligned}
f&=(1-A^2r^2)f_0+\frac{r^2}{\ell^2}\,,\nonumber\\
h&=1+2Am\cos\theta+A^2w^2\cos^2\theta\,,\nonumber\\
\Omega&=1+Ar\cos\theta,\qquad w^2=e^{-\gamma}(q_e^2+q_m^2),
\end{aligned}
\end{equation}
being $f_0$ the static metric function characterizing the (asymptotically flat) static solution
\begin{equation}
f_0=1-\frac{2m}{r}+\frac{w^2}{r^2}\,,  \label{f00}
\end{equation}
together with the corresponding ModMax gauge field
\begin{equation}
\mathcal{A}=-\frac{e^{-\gamma}q_e }{ r}dt+q_m\cos\theta \frac{d\varphi}{K},\qquad \mathcal{F}=d\mathcal{A}. 
\end{equation}
The solution is characterized by the five parameters $A,\,m,\,q_e,\,q_m$ and $K$, being respectively, the acceleration, the mass, the electric and magnetic charges, and the conical deficits. An appropriate choice of $K$ can remove one of the spacetime's conical singularities; we will return to this point later. Note that the effect of the nonlinearity of ModMax consists of ``screening'' the charges by the factor of $e^{-\gamma}$.

In the next section, we will make use of the original method proposed by Havrdova and Krtou{\v s} \cite{Havrdova:2006gi, Havrdovathesis} to land in the corresponding ModMax electromagnetic universe via the C-metric configuration \eqref{Cmetricmodmax}. In order to achieve this task, it is more convenient to express the above solution (without regard to the cosmological constant for now) in prolate coordinates $(x,y)$. Indeed, performing a change of coordinates of the form
\begin{equation}
t=\frac{\bar{t}}{A},\quad r=\frac{1}{Ay},\quad \cos\theta=x,
\end{equation}
the line element reads
\begin{equation}\label{cmetricprolate}
ds^2=\frac{1}{A^2(x+y)^2}\left[-Fd\bar{t}^2+\frac{dy^2}{F}+\frac{dx^2}{G}+G\frac{d\varphi^2}{K^2}\right],
\end{equation}
where
\begin{equation}
\begin{aligned}\label{cmpolinomials}
F(y)&=-(1-y^2)(1-2Amy+A^2w^2y^2),\\ G(x)&=(1-x^2)(1+2Amx+A^2w^2x^2),
\end{aligned}
\end{equation}
while the electromagnetic tensor solution is given by
\begin{equation}\label{modmaxtensor}
\mathcal{F}=-e^{-\gamma}q_e  dy\wedge d\bar{t}+q_m dx\wedge \frac{d\varphi}{K}.
\end{equation}
The spacetime coordinates are restricted as $t\in\mathbb{R}$, $x\in(-1,1)$, $y<-x$ and $\varphi\in(-\pi,\pi)$. In this convenient set of coordinates originally presented in \cite{Hong:2003gx}, it is easy to observe that $F(z)=-G(-z)$, which means that these two functions will have the same number of roots. In fact, the roots of $F(y)$ are explicitly given by 
\begin{equation}
\begin{aligned}
   & y_c=-1,\quad y_A=1,\quad y_{i}=\frac{1}{Aw^2}(m-\sqrt{m^2-w^2}),\\
   & y_{o}=\frac{1}{Aw^2}(m+\sqrt{m^2-w^2}), 
\end{aligned}
\end{equation}
where $y_A$ represents the accelerating horizon, while $y_{i}$ and $y_{o}$ are the inner and outer black hole horizons naturally contained in a charged configuration. Notice that $y_c=-1$ lies outside the physical causal structure of the solution. 
As usual in a C-metric configuration with no external background fields, the acceleration mechanism is provided by the occurrence of conical singularities along each of the semi-axis
\begin{equation}
    \delta_{x=\pm1}=2\pi\left(1-\frac{|1\pm2Am+A^2w^2|}{|K|}\right).
\end{equation}
In what follows, we fix the conicity parameter to be $K=1+2Am+A^2w^2$, in such a manner that the conical deficit along the positive semi-axis is removed. The whole conicity is translated to the negative semi-axis $x=-1$.

\section{The Melvin-Bonnor-ModMax spacetime}\label{secIII}

After presenting the C-metric configuration in ModMax \eqref{Cmetricmodmax}, we proceed with the construction of the Melvin-Bonnor-ModMax spacetime via the limiting procedure exposed in \cite{Havrdova:2006gi, Havrdovathesis}. In essence, this limit corresponds to a self-gravitating extension of what is known for the case of a homogeneous test electric field in Minkowski spacetime. In such a situation, the homogeneous electric field is achieved starting from the configuration of two charged accelerating particles by enlarging the distance between them, while properly increasing the value of their charge. In the backreacting extension, it is then direct to consider the utilization of the spacetime of two charged accelerating black holes, namely the C-metric configuration \cite{Barrientos:2022bzm}. The limiting procedure described in \cite{Havrdova:2006gi, Havrdovathesis} consists of two main ingredients: a wisely defined coordinate transformation that allows performing a near horizon approximation around the Rindler horizon of the accelerating geometry and a proper redefinition of the black hole inner and outer horizons, redefinition that ultimately prescribed the behavior of the physical parameters $m$ and $w$ such that the limit as a solution of the Einstein-ModMax field equations is guaranteed. The parameter $A$ is fixed during the limiting process.  

In mathematical terms, the inner and outer horizons are reparametrized as follows
\begin{equation}
    y_{i}=1+\tilde{y}_{i}\epsilon,\quad  y_{o}=1+\tilde{y}_{o}\epsilon \label{newhorizons}, 
\end{equation}
being $\tilde{y}_i$ and $\tilde{y}_0$ two constants related to $m$, $w$ and $A$ and satisfying $\tilde{y}_i<\tilde{y}_o$; this for the order of the horizons to be maintained. The parameter $\epsilon$ is a small parameter to be sent to zero when performing the limit. For the limit to be carried around the Rindler horizon, the following change of coordinates has to be considered
\begin{equation}
    \bar{t}=\frac{1}{\tilde{y}_i\tilde{y}_o}\frac{1}{\epsilon^2}\tau,\quad y=1+\tilde{y}_i\tilde{y}_o\epsilon^2 v. \label{newcoordiantes}
\end{equation}
The coordinates $x$ and $\varphi$ remain unchanged. Notice that this implies that $y_i$ and $y_o$ scale as $\epsilon$ causing the new horizon locations $v_i$ and $v_o$ to scale as $\epsilon^{-1}$. As a consequence, the horizons are pushed away in the limit $\epsilon\rightarrow0$, confirming the analogy with the flat spacetime case in which the charges are pushed away from each other, leaving a remnant electric field between them \cite{Havrdova:2006gi, Havrdovathesis}. In addition, the relevant causal structure condition $y<-x$ becomes  $v<\infty$, while the three roots of the function $G$ degenerate to $x=-1$ in the limit. As a result of the simultaneous consideration of $\eqref{newhorizons}$ and $\eqref{newcoordiantes}$, in the limit $\epsilon\rightarrow0$, the full configuration \eqref{Cmetricmodmax} takes the form
\begin{eqnarray}
\begin{aligned}\label{firstmelvin}
    ds^2&=\frac{1}{A^{2}(1+x)^2}\left[-2vd\tau^2+\frac{dv^2}{2v}+\frac{dx^2}{(1-x)(1+x)^3}\right.\\ 
    &\quad\left.+\frac{(1-x^2)d\varphi^2}{16A^2}\right],\\
    \mathcal{F}&=-e^{-\gamma}q_e dv \wedge d\tau+\frac{q_m}{4}dx\wedge d\varphi.   
\end{aligned}
\end{eqnarray}
Notice that \eqref{newhorizons} implies that $(m,w)\rightarrow1/A$ in the limit of vanishing $\epsilon$, a necessary condition for \eqref{firstmelvin} to solve the Einstein-ModMax field equations. The spacetime configuration \eqref{firstmelvin} already represents the line element and gauge field profile of the Melvin-Bonnor-ModMax configuration; however, in Rindler-like coordinates, the coordinates adapted to an accelerating observer. To properly identify the Melvin-Bonnor-ModMax universe in an intuitive form, we move to the coordinates of a static global observer as suggested in Refs. \cite{Havrdova:2006gi, Havrdovathesis}. We start with a suitable redefinition of the acceleration parameter 
\begin{equation}
    A=\frac{e^{-\gamma/2}\sqrt{E^2+B^2}}{4}, 
\end{equation}
and the following  change of coordinates 
\begin{eqnarray}
\begin{aligned}
    v&=\frac{e^{-\gamma}(E^2+B^2)}{8}(-t^2+z^2),\quad \tau=\tanh^{-1}{\left(\frac{t}{z}\right)},\\
    x&=\frac{1-\frac{e^{-\gamma}(E^2+B^2)}{4}\rho^2}{1+\frac{e^{-\gamma}(E^2+B^2)}{4}\rho^2}.
\end{aligned}
\end{eqnarray}
As a result, the line element of the Melvin-Bonnor-ModMax spacetime, in cylindrical coordinates, takes the recognizable form 
\begin{equation}
\begin{aligned}
ds^2&=\left[1+\frac{e^{-\gamma}(E^2+B^2)}{4}{\rho}^2\right]^2(-d{t}^2+d{\rho}^2+d{z}^2)\\
&\quad+\frac{{\rho}^2}{\left[1+\frac{e^{-\gamma}(E^2+B^2)}{4}{\rho}^2\right]^2}d\varphi^2,
\end{aligned}
\end{equation}
while its gauge field, via the ulterior redefinition 
\begin{eqnarray}
    q_e=\frac{4E}{e^{-\gamma}(E^2+B^2)},\quad q_m=\frac{4B}{e^{-\gamma}(E^2+B^2)},
\end{eqnarray}
becomes
\begin{equation}
\mathcal{F}=e^{-\gamma}E d{z}\wedge d{t}+\frac{B{\rho}}{\left[1+\frac{e^{-\gamma}(E^2+B^2)}{4}{\rho}^2\right]^2} d{\rho}\wedge d\varphi.\label{melvinMODMAXgauge}
\end{equation}
Note that, here,  the time coordinate $t$ must not be confused with the original time coordinate used in \eqref{Cmetricmodmax}. 

This configuration describes the backreaction of a parallel bundle of electromagnetic flux in ModMax. As we can observe, the electric and magnetic fields are screened by a factor $e^{-\gamma}$; however, the causal structure of the solution is practically unchanged with respect to the Melvin-Bonnor spacetime in Einstein-Maxwell. It belongs to the Kundt class of type D spacetimes and asymptotically approaches the Levi-Civita spacetime, modulo a reparametrization of the noncompact coordinates \cite{Griffiths:2009dfa}.  
The main features of this spacetime have been extensively studied in the literature \cite{Griffiths:2009dfa, PhysRev.139.B244, melvin1966}, and hence it is unnecessary to reproduce them here. Notwithstanding, it is important to stress that the solution remains nonsingular, as the electromagnetic theory under consideration is not Maxwell's theory anymore. For a generic NLE the occurrence of curvature singularities will depend on the form of the NLE Lagrangian \cite{Gibbons:2001sx}, usually imposing bounds on $E$ and $B$ for the solution to be curvature singularity free. This is not the case with ModMax, where a straightforward exploration of the curvature scalars, the Kretschmann invariant $\mathcal{K}:=R_{\mu\nu\rho\sigma}R^{\mu\nu\rho\sigma}$, reveals the singularity free nature of the spacetime
\begin{equation}\label{kretchmann}
\mathcal{K}=\frac{\left[\frac{3(E^2+B^2)^2\rho^4}{e^{2\gamma}}-\frac{24(E^2+B^2)\rho^2}{e^{\gamma}}+80\right](E^2+B^2)^2}{4e^{2\gamma}\left(1+\frac{e^{-\gamma}(E^2+B^2)}{4}\rho^2\right)^8}.
\end{equation}
In addition, the charges of the solution are proven to be finite, just as in the case of the Einstein-Maxwell theory
{
\begin{equation}
\begin{aligned}
\mathcal{Q}_e&=\frac{1}{4\pi}\int_\mathcal{H} *\mathbf{E} = \frac{e^{\gamma}E}{E^2+B^2},\\
 \mathcal{Q}_m&=\frac{1}{4\pi}\int_\mathcal{H} \mathcal{F} = \frac{e^{\gamma}B}{E^2+B^2}.
\end{aligned}
\end{equation}
}
This agrees with the gravitational bounded nature of the homogeneous electromagnetic field of the Melvin-Bonnor-ModMax configuration and once again contrasts what occurs in the flat spacetime case. 

In Einstein-Maxwell, the limit proposed in \cite{Havrdova:2006gi, Havrdovathesis} has been applied to the case in which a cosmological constant is included \cite{Lim:2018vbq}. The procedure applies in direct analogy with respect to the asymptotically flat case starting from the factorized form of the charged (anti)-de Sitter C-metric proposed in \cite{Chen:2015vma}. Here, we extend those findings to the case of Einstein-ModMax and deliver the (A)dS Melvin-Bonnor-ModMax spacetime in a suitable set of coordinates in which the $\Lambda=0$ case is easily retrieved \cite{Barrientos:2024pkt}. The configuration solution, in this case, reads 
\vspace{-1cm}
\begin{widetext}
\begin{equation*}
\begin{aligned}
ds^2&=\left[1+\frac{e^{-\gamma}(E^2+B^2)}{4}\rho^2\right]^2\left(-dt^2+\frac{\rho^2d\rho^2}{\rho^2-\frac{4}{3}\frac{\Lambda}{e^{-2\gamma}(E^2+B^2)^2}\left[1+\frac{e^{-\gamma}(E^2+B^2)}{4}\rho^2\right]^4}+dz^2\right)\\
&\quad +\frac{\rho^2-\frac{4}{3}\frac{\Lambda}{e^{-2\gamma}(E^2+B^2)^2}\left[1+\frac{e^{-\gamma}(E^2+B^2)}{4}\rho^2\right]^4} {\left[1+\frac{e^{-\gamma}(E^2+B^2)}{4}\rho^2\right]^2}d\varphi^2,
\end{aligned}
\end{equation*}
\end{widetext}
\begin{equation}
\mathcal{A}=e^{-\gamma}Ez dt+\frac{B\rho^2}{2\left[1+\frac{e^{-\gamma}(E^2+B^2)}{4}\rho^2\right]}d\varphi.
\end{equation}
As noted in \cite{Barrientos:2024pkt}, the spacetime in question involves the presence of a spinning string. At a fixed $z$, we have 
\begin{equation}
\lim_{\rho\rightarrow 0}g_{\varphi\varphi}=-\frac{4}{3}\frac{e^{2\gamma}\Lambda}{(E^2+B^2)^2}.
\end{equation}
When the cosmological constant is positive, removing the string is impossible; the parameter $\gamma$ can not be used to tune the constants to any particular value such that the string is removed. Consequently, the de-Sitter electromagnetic spacetime also inherits the cosmic string from the C-metric \cite{Lim:2018vbq}. However, if $\Lambda < 0$, we can reparametrize the coordinates as
\begin{equation}
(t,\varphi)=\left(\tilde{t}+\frac{2\sqrt{-\Lambda}}{\sqrt{3}e^{-\gamma}(E^2+B^2)}\tilde{\varphi},\tilde{\varphi}\right),
\end{equation}
and re-glue the spacetime to eliminate the string. The resulting spacetime is then free of conical singularities. Choosing a negative cosmological constant also ensures the metric retains a Lorentzian signature. By analyzing the Kretschmann scalar, we can further confirm that the spacetime is free of curvature singularities for any value of the cosmological constant
\begin{equation}
R_{\mu\nu\rho\sigma}R^{\mu\nu\rho\sigma}=\frac{8\Lambda}{3}+\mathcal{K}_0,
\end{equation}
where $\mathcal{K}_0$ corresponds to the value of the Kretschmann scalar given in equation \eqref{kretchmann}.

\section{Static black holes within the Melvin-Bonnor-ModMax Universe}\label{secIV}

It is well-known that the construction of magnetized black hole solutions \cite{Ernst:1976mzr, 10.1063/1.522935} of the Einstein-Maxwell equations has been made possible by exploiting the Lie point symmetries of the field equations \cite{Ernst:1967wx, Ernst:1967by}. This approach leverages the symmetries inherent in the equations in order to generate new solutions from known ones. However, this method is highly dependent on the specific form of the underlying theory and cannot be easily adapted to accommodate even small modifications of the action. To illustrate this, even the simple addition of a cosmological constant to the action complicates the generalization of the method. This constraint is the reason why we use limiting procedures to construct the Melvin-(A)dS solution \cite{Gibbons:2001sx, Lim:2018vbq, Barrientos:2024pkt}. The challenges are even more pronounced when considering nonlinear electrodynamics. In these cases, the nonlinearity of the field equations breaks the symmetries that the Ernst mechanism relies on, rendering this method completely inapplicable. Consequently, constructing (electro)magnetized black holes within NLE models becomes a highly nontrivial task. Surprisingly, due to the ``Maxwell-like'' character of static solutions in ModMax, we can embed static black hole solutions within the Melvin-Bonnor-ModMax universe. Specifically, we construct Schwarzschild and accelerating black holes embedded in the ModMax electromagnetic geometry.

The key insights for the existence of such a configuration are twofold. First, it is observed in Einstein-Maxwell theory that the presence of a Schwarzschild black hole onto the Melvin-Bonnor spacetime modifies the form of the electromagnetic field in a very precise manner, that it does not alter the magnetic field while introducing the lapse function of the line element into the electric component. Second, the Melvin-Bonnor-ModMax spacetime, just as the C-metric solution \cite{Barrientos:2022bzm} and all spherically symmetric configurations in ModMax satisfy the key condition $\mathcal{S}\propto\mathcal{P}$, in fact, for the Melvin-Bonnor-ModMax configuration we have 
\begin{equation}
\begin{aligned}
\mathcal{S}&=\frac{B^2-e^{-2\gamma}E^2}{\left(1+\frac{e^{-\gamma}(E^2+B^2)}{4}\rho^2\right)^4},\\ \mathcal{P}&=-\frac{2e^{-\gamma}EB}{\left(1+\frac{e^{-\gamma}(E^2+B^2)}{4}\rho^2\right)^4}.
\end{aligned}
\end{equation}
This enormously reduces the field equations overlapping the spectrum of solutions in Einstein-Maxwell and Einstein-ModMax theories modulo a constant factor to be redefined. 

Hence, the Schwarzschild-Melvin-Bonnor-ModMax black hole takes the simple form 
\begin{widetext}
\begin{equation}\label{schMelvinMODMAX}
\begin{aligned}
    ds^2&=\left[1+\frac{e^{-\gamma}(E^2+B^2)}{4}r^2\sin^2\theta\right]^2\left(-fdt^2+\frac{dr^2}{f}+r^2d\theta^2\right)+\frac{r^2\sin^2\theta}{\left[1+\frac{e^{-\gamma}(E^2+B^2)}{4}r^2\sin^2\theta\right]^2}d\varphi^2,\\
\mathcal{A}&=e^{-\gamma}Erf\cos\theta dt+\frac{Br^2\sin^2\theta}{2\left[1+\frac{e^{-\gamma}(E^2+B^2)}{4}r^2\sin^2\theta\right]}d\varphi,\quad f=1-\frac{2m}{r}.
\end{aligned}
\end{equation}
\end{widetext}
Notice that we have written the metric in spherical-like coordinates to favor comparison with the Schwarzschild solution. In this case, we explicitly obtain 
\begin{align}
\mathcal{S}&=\frac{\left(f+\frac{2m}{r}\cos^2\theta\right)\left(B^2-e^{-2\gamma}E^2\right)}{\left(1+\frac{e^{-\gamma}(E^2+B^2)}{4}r^2\sin^2\theta\right)^4},\nonumber\\
\mathcal{P}&=-\frac{2\left(f+\frac{2m}{r}\cos^2\theta\right)e^{-\gamma}EB}{\left(1+\frac{e^{-\gamma}(E^2+B^2)}{4}r^2\sin^2\theta\right)^4}.
\end{align}

The condition $\mathcal{S}\propto\mathcal{P}$ can be further explored to integrate even more intricate black hole spacetimes within the Melvin-Bonnor-ModMax background; in fact, it is possible to construct the accelerating extension of \eqref{schMelvinMODMAX}, the C-metric-Melvin-Bonnor-ModMax black hole configuration
\begin{widetext}
    \begin{equation}
\begin{aligned}
ds^2&=\frac{1}{\Omega^2}\left(\left[1+\frac{e^{-\gamma}(E^2+B^2)}{4}\frac{Pr^2\sin^2\theta}{\Omega^2}\right]^2\left(-Qdt^2+\frac{dr^2}{Q}+\frac{r^2 d\theta^2}{P}\right)+\frac{Pr^2\sin^2\theta}{\left[1+\frac{e^{-\gamma}(E^2+B^2)}{4}\frac{Pr^2\sin^2\theta}{\Omega^2}\right]^2}d\varphi^2\right),\\
\mathcal{A}&=e^{-\gamma}Erf\frac{2\cos\theta+Ar(1+\cos^2\theta)}{2\Omega^2} dt+\frac{BPr^2\sin^2\theta}{2\Omega^2\left[1+\frac{e^{-\gamma}(E^2+B^2)}{4}\frac{Pr^2\sin^2\theta}{\Omega^2}\right]}d\varphi,
\end{aligned}
\end{equation}
\end{widetext}
where for simplicity we have defined $Q=\left(1-A^2r^2\right)f$, $P=1+2Am\cos\theta$ and $\Omega=1+Ar\cos\theta$, with $f$ defined in \eqref{schMelvinMODMAX}. The invariants explicitly read
\begin{equation}
\begin{aligned}
\mathcal{S}&=\frac{\left((P^2+A^2m^2\sin^4\theta)r-2Pm\sin^2\theta\right)\left(B^2-e^{-2\gamma}E^2\right)}{r\left(1+\frac{e^{-\gamma}(E^2+B^2)}{4}\frac{Pr^2\sin^2\theta}{\Omega^2}\right)^4},\\
\mathcal{P}&=-\frac{2\left((P^2+A^2m^2\sin^4\theta)r-2Pm\sin^2\theta\right)e^{-\gamma}EB}{r\left(1+\frac{e^{-\gamma}(E^2+B^2)}{4}\frac{Pr^2\sin^2\theta}{\Omega^2}\right)^4}.
\end{aligned}
\end{equation}
Notice that in all the cases we have exposed, namely, the Melvin-Bonnor-ModMax and its Schwarzschild and C-metric generalizations, the ratio $\mathcal{S}/\mathcal{P}$ acquires the same value. These black hole spacetimes represent the first
examples of a black hole ever embedded in an electromagnetic background for an NLE model. 

\subsection{Novel Kerr-Schild representations}

As discussed earlier, the geometric properties and causal structures of the spacetimes considered in this work have been extensively examined in the literature \cite{Griffiths:2009dfa, Barrientos:2024pkt, Ernst:1976mzr, 10.1063/1.522935, Stephani:2003tm}. For this reason, we refrain from delving into those aspects further. However, we offer two novel and, to the best of our knowledge, intriguing observations concerning unknown Kerr-Schild representations of the Schwarzschild-Melvin-Bonnor and C-metric spacetimes. While our constructions are presented explicitly for these geometries within the framework of Einstein-ModMax theory, they remain equally applicable in the context of electrovacuum.

We begin by showing that the Schwarzschild-Melvin-Bonnor-ModMax black hole solution \eqref{schMelvinMODMAX} can be cast in a Kerr-Schild form. As is well known, the Kerr-Schild construction allows mass to be introduced from an initially massless seed configuration. Typically, for charged solutions such as Reissner-Nordström, Kerr-Newman, or even those of dilatonic origin, the potential vector does not depend on the mass. Consequently, the Kerr-Schild transformation only acts on the metric and not on the electromagnetic field. However, in the solution \eqref{schMelvinMODMAX}, the gauge field explicitly depends on the black hole's mass, necessitating a generalization of the Kerr-Schild transformation to account for this dependence. Here, we present for the first time a Kerr-Schild construction of this black hole solution. As the construction is novel, we will present it step by step for clarity. Our starting point is the seed solution, hereafter denoted with a $0$ subscript,  
\begin{eqnarray}\label{schMelvinMODMAXseed}
\begin{aligned}
    ds_0^2&=H(r,\theta)^2\left(-dt^2+dr^2+r^2d\theta^2\right)+\frac{r^2\sin^2\theta}{H(r,\theta)^2}d\varphi^2,\\
\mathcal{A}_0&=e^{-\gamma}Er\cos\theta dt+\frac{Br^2\sin^2\theta}{2 H(r,\theta)}d\varphi,\\
\end{aligned}
\end{eqnarray}
where for simplicity we have defined the function 
$$
H(r,\theta)=1+\frac{e^{-\gamma}(E^2+B^2)}{4}r^2\sin^2\theta.
$$
It is simple to see that a null, geodesic, and shear-free one-form field for this metric is given by $l=dt-dr$. Hence, let us consider the following Kerr-Schild ansatz
\begin{subequations}\label{melvinKS}
\begin{eqnarray}\label{gmelvinKS}
ds^2&=&ds_0^2+Z(r,\theta)\left(dt-dr\right)^2,\\
\label{FmelvinKS}
\mathcal{A}&=&\mathcal{A}_0+A_1(r,\theta)\left(dt-dr\right).
\end{eqnarray}
\end{subequations}
The novelty compared to known static charged solutions as in the Reissner-Nordstrom case is that the gauge field is decomposed as a linear combination of a seed potential $\mathcal{A}_0$ and a piece proportional to the null vector $l$. Moreover, as shown below, this latter piece explicitly depends on the mass. For instance, even in the rotating Kerr-Newmann case, one can see that the Maxwell $U(1)-$field, in the Kerr-Schild derivation has no seed contribution, and although is proportional to the null, geodesic and shearfree congruence \cite{Ayon-Beato:2015nvz}, this factor of proportionality only depends on the electric charge.

Starting from such new ansatz \eqref{melvinKS}, and denoting the Einstein-ModMax equations by ${\cal E}_{\mu\nu}$ and ${\cal M}^{\mu}$, one can note that the equation ${\cal E}_{\theta\phi}=0$ imposes 
\begin{equation}
Z(r,\theta)=-\frac{\partial_\theta A_1(r,\theta)}{Er\sin\theta} H(r,\theta)^2,
\end{equation}
and that the Maxwell equation ${\cal M}^{\varphi}=0$ imposes $A_1(r,\theta)$ to be as a separable sum, i.e., $A_1(r,\theta)=F_1(r)+F_2(\theta)$. Additionally, due to the equation ${\cal E}_{t \varphi}=0$, one ends up with   \begin{subequations}\label{melvinKS1}
\begin{eqnarray}\label{gmelvinKS1}
ds^2&=&ds_0^2+\frac{2m}{r}H(r,\theta)^2\left(dt-dr\right)^2,\\
\label{FmelvinKS1}
\mathcal{A}&=&\mathcal{A}_0+2mE\cos\theta\left(dt-dr\right)\\
&=&Er\cos\theta\left(1-\frac{2m}{r}\right)dt-2mE\cos\theta dr\nonumber\\
&\qquad+&\frac{Br^2\sin^2\theta}{2H(r,\theta)}d\varphi.
\end{eqnarray}
\end{subequations}
As usual, the off-diagonal terms of the Kerr-Schild metric \eqref{gmelvinKS1} can be canceled by a change of variable, and hence one yields precisely configuration \eqref{schMelvinMODMAX}. Additionally, 
it is also interesting to note that the Kerr-Schild transformation \eqref{melvinKS1} has the following remarkable representation
\begin{equation}
\begin{aligned} g_{\mu\nu}&=g^{(0)}_{\mu\nu}-\left(\frac{2m}{r}\right) \left[g^{(0)}_{ \alpha\beta}\, \xi_{(0)}^{\alpha}\xi_{(0)}^{\beta}\right]  \,l_{\mu}\otimes l_{\nu},\\
   \mathcal{A}_{\mu}&={\cal A}^{(0)}_{\mu}-\left(\frac{2m}{r}\right)\left[ {\cal A}^{(0)}_{\alpha}\,\xi_{(0)}^{\alpha} \right]\,l_{\mu},
\end{aligned}
\end{equation}
where $\xi_{(0)}^{\alpha}\partial_{\alpha}=-\partial_t$ is a timelike Killing vector field.

Continuing with our new Kerr-Schild approach, one can show that the C-metric solution \eqref{Cmetricmodmax}, as well as the standard C-metric of Einstein-Maxwell, has also a generalized Kerr-Schild representation, which, to our knowledge, has never been put in light before. As in the previous case, since this derivation is original, we intend to detail each step. Nevertheless, let us re-derive the solution \eqref{Cmetricmodmax} only in the purely electric case and without a cosmological constant. In fact, the cosmological constant can be easily incorporated into the seed metric, while the magnetic contribution of the gauge field can be added later by applying duality, as is typically done. 

We start from the following seed configuration,
\begin{eqnarray}
\begin{aligned}\label{Cmetricmodmax2}
ds_0^2&=\frac{1}{\Omega^2}\Bigl(-f_0dt^2+\frac{dr^2}{f_0}+r^2\Bigr[\frac{d\theta^2}{h_0}-h_0\sin^2\!\theta \frac{d\varphi^2}{K^2}\Bigr]\Bigr),\\
\mathcal{A}&=-\frac{e^{-\gamma}q_e }{ r}dt. 
\end{aligned}
\end{eqnarray}
where we have defined
\begin{equation}
\begin{aligned}
f_0&=(1-A^2r^2)(1+\frac{w^2}{r^2})\,,\nonumber\\
h_0&=1+A^2w^2\cos^2\theta\,,\nonumber\\
\Omega&=1+Ar\cos\theta,\qquad w^2=e^{-\gamma}q_e^2.
\end{aligned}
\end{equation}
This definition of $f_0$ should not be confused with the one provided in equation \eqref{f00}. It is important to note that the seed metric is a solution with signature $\mbox{sgn}(2,2)$ since the metric component $g_{(0)\varphi\varphi}<0$. Hence, let us operate a double Kerr-Schild transformation on the $\mbox{sgn}(2,2)$ seed metric given by 
\begin{equation}
\label{kkkk}
ds^2=ds_0^2+\frac{H_1(r)}{\Omega^2}l\otimes l+\frac{r^2H_2(\theta)}{\Omega^2} k\otimes k,
\end{equation}
where we have defined 
\begin{equation}
\label{H1H2}
H_1(r)=\frac{2m}{r}(1-A^2r^2),\quad H_2(\theta)=-\frac{2mA\cos\theta}{h_0^2},
\end{equation}
and, where the null vectors $l$ and $k$ along which the Kerr-Schild transformations are operated are given by
\begin{eqnarray}
    l=dt-\frac{dr}{f_0},\qquad k=d\theta+\frac{h_0\sin\theta}{K}d\varphi.
\end{eqnarray}
Note that $l$, as usual, is geodesic and shear-free. The novelty here is that, although $k$ is not geodesic, $k^a\left(\nabla_a k_{b}\right) \neq 0$, the norm of the geodesic vector is unexpectedly zero, 
\begin{eqnarray}
k^a\left(\nabla_a k_{b}\right)k^c\left(\nabla_c k^b\right)=0.
\label{normgeo}
\end{eqnarray}
Finally, by means of the coordinate transformations
\begin{eqnarray*}
dt&\to &dt+\frac{H_1(r)}{f_0(r)\left(H_1(r)-f_0(r)\right)}dr,\\
d\varphi&\to& d\varphi-\frac{K H_2(\theta)}{\left(H_2(\theta)h_0(\theta)-1\right)\sin\theta}d\theta,
\end{eqnarray*}
that eliminate off-diagonal terms, and after the Wick rotation $\varphi\to i\varphi$ in \eqref{kkkk}, the metric \eqref{Cmetricmodmax} is recovered. 

In summarize, we have highlighted unexpected aspects of the Schwarzschild-Melvin-Bonnor-ModMax solutions \eqref{schMelvinMODMAX}, as well as the C-metric \eqref{Cmetricmodmax}, by showing that each of these solutions admits a novel Kerr-Schild type representation.

\section{A vortex-like background in ModMax}\label{secV}

To construct our swirling spacetime in ModMax, it is instructive, as a prelude, to revisit the case of the Taub-NUT geometries already existing in the literature \cite{BallonBordo:2020jtw, Flores-Alfonso:2020nnd}. Contrary to the accelerating solutions of the previous sections, the Taub-NUT spacetimes, as we will also see for the swirling case, the condition $\mathcal{S} \propto \mathcal{P}$ is not respected, and thus we can expect a more significant deviation from Einstein-Maxwell configurations. However, $\mathcal{S}$ and $\mathcal{P}$ behave in such a way that the integrability of the ModMax field equations is not compromised.

The Taub-NUT solutions are described by the line element 
\begin{equation} ds^2 = -f(r)(dt - 2n\cos\theta  d\varphi)^2 + \frac{dr^2}{f(r)} + (r^2 + n^2) d\Omega^2, 
\end{equation} 
where, as usual, $d\Omega^2 = d\theta^2 + \sin^2\theta  d\varphi^2$ represents the line element of the 2-sphere, and $n$ denotes the NUT parameter. Solutions have been found for the simplest choice of gauge field, specifically one compatible with the underlying $U(1)$ Hopf fibration of the spacetime, in which the standard monopole contribution proportional to $q_m\cos\theta$ is not explicitly visible.

Here, we write this solution in the more convenient gauge 
\begin{equation} 
\mathcal{A} = A_t(r)(dt - 2n\cos\theta  d\varphi) + q_m\cos\theta d\varphi. \label{gaugenostro} 
\end{equation} 
The Taub-NUT spacetime in ModMax then takes the form
\begin{widetext}
\begin{equation}
\begin{aligned}
    ds^2&=-\left(\frac{r^2-2mr-n^2+e^{-\gamma}(q_e^2+q_m^2)}{r^2+n^2}\right)(dt-2n\cos\theta d\varphi)^2+\frac{dr^2}{\left(\frac{r^2-2mr-n^2+e^{-\gamma}(q_e^2+q_m^2)}{r^2+n^2}\right)}+(r^2+n^2)d\Omega^2,  \\
    A_t&=  - \frac{q_e}{2n}\sin\left(e^{-\gamma}\left[\pi-2\arctan\left(\frac{r}{n}\right)\right]\right)
    -\frac{q_m}{2n}\cos\left(e^{-\gamma}\left[\pi-2\arctan\left(\frac{r}{n}\right)\right]\right)+ \frac{q_m}{2n}.  \label{ourTaub-NUT}
\end{aligned}  
\end{equation}
\end{widetext}
Following the same procedures outlined in \cite{BallonBordo:2020jtw, Flores-Alfonso:2020nnd}, it is straightforward to verify the limits of \eqref{ourTaub-NUT} to its standard Einstein-Maxwell form \cite{Barrientos:2024pkt} and to the dyonic Reissner-Nordstr\"om solution in ModMax \cite{Flores-Alfonso:2020euz, Amirabi:2020mzv}. The existence of these solutions hinges on how the ModMax interaction specifically modifies the Maxwell equations. Having the following electromagnetic invariants
\begin{equation}
    \mathcal{S}=-(A^{\prime}_t)^2+\left(\frac{2nA_t-q_m}{r^2+n^2}\right)^2,\quad\mathcal{P}=\frac{2A^{\prime}_t(2nA_t-q_m)}{r^2+n^2}
\end{equation}
the ModMax equation modifies the standard Maxwell equation by introducing different constant pre-factors in front of each term of the Maxwell differential equation to be solved. However, the fundamental structure of the equation remains unchanged
\begin{equation}
   e^{\gamma} A^{\prime\prime}_t+e^{\gamma}\frac{2rA^{\prime}_t}{r^2+n^2}+e^{-\gamma}\frac{2n(2nA_t-q_m)}{(r^2+n^2)^2}=0,
\end{equation}
up to the exponentials of the $\gamma$ parameter. 
The overall integration of the ModMax equation is therefore not compromised, but the form of the gauge potential is slightly modified, yielding solution \eqref{ourTaub-NUT}. Due to the presence of $e^{-\gamma}$
inside the trigonometric sine and cosine functions, this solution does not reduce to the usual form in Einstein-Maxwell theory. The key point behind this simplification is the form of the gauge potential, particularly the term proportional to $\omega A_t$, the induced magnetic term due to the stationarity of the spacetime.

To broaden the spectrum of solutions in Einstein-ModMax theory, it is wise to adopt a similar approach and seek an electromagnetic configuration where an induced field of the previous form is achieved. We can draw on insights from the Ernst approach to Einstein-Maxwell theory \cite{Ernst:1967wx, Ernst:1967by}. It is known that the Reissner-Nordstr\"om-NUT spacetime can be obtained, in addition to direct integration, through the successive application of an electric Harrison transformation \cite{harrison1968new}, which transforms a Schwarzschild spacetime into a Reissner-Nordstr\"om configuration, followed by an electric Ehlers transformation that ultimately introduces the NUT parameter \cite{Ehlers:1957zz, Ehlers:1959aug}.
Due to the structure of these electric transformations, the so-called twisted Ernst equations—integrability conditions of the Einstein-Maxwell system expressed in terms of complex Ernst potentials—naturally induce a magnetic gauge field component of the desired form, specifically proportional to $\omega A_t$.
The most promising approach to introduce a new stationary spacetime in Einstein-ModMax with the desired electromagnetic configuration is to consider a Melvin-Bonnor spacetime endowed with a vortex-like rotation \cite{Barrientos:2024pkt, Capobianco:2024jhe}. A spacetime closely related to the Taub-NUT spacetime is the swirling or vortex-like spacetime \cite{Gibbons:2013yq, Astorino:2022aam,Capobianco:2023kse}. Unlike the standard Kerr-like rotation, where the spacetime itself does not rotate unless a rotating source induces nontrivial dragging, the swirling geometry features a background that rotates and drags bodies along with it, rather than being dragged by a source. 
Given that mixing electric and magnetic Ehlers or Harrison transformations disrupts the commutativity of the process, and after analyzing these backgrounds in \cite{Barrientos:2024pkt}, an educated guess guides us to consider a Melvin-Bonnor-Swirling spacetime, which in the Ernst scheme is constructed by composing a Harrison transformation with an Ehlers transformation of the magnetic type, thereby providing a gauge field with an induced electric field of the desired form via the corresponding twisted equations, $\sim\omega A_{\varphi}$.
With this approach, the integration of the Einstein-ModMax equations remains intact, just as with the Taub-NUT case. The relevant Maxwell equation, specifically the magnetic component (noting that the swirling spacetime can be viewed as a variant of the Taub-NUT geometry achieved through a magnetic Ehlers transformation), differs from the Einstein-Maxwell case only in the pre-factors in front of each term of the differential equation. In this case, the equation reads
\begin{equation}
\begin{aligned}
e^{-\gamma}A_\varphi^{\prime\prime}+\left(2VV^\prime-\frac{V^2-3j^2\rho^4}{\rho}\right)\frac{e^{-\gamma}A_\varphi^\prime}{(V^2+j^2\rho^4)}\\
+\frac{4j(E+4je^{\gamma}A_\varphi)\rho^2}{(V^2+j^2\rho^4)^2}=0.
\end{aligned}
\end{equation}
See below for the ansatz of the line element and gauge field, Eq. \eqref{swirlingmelvinansatz}. Notice that, as well as in the Taub-NUT case, the electromagnetic invariants $\mathcal{S}$ and $\mathcal{P}$ are not proportional to each other
\begin{equation}
\begin{aligned}
\mathcal{S}&=\frac{A_\varphi ^{\prime 2}}{\rho^2}-\frac{\left(4j A_\varphi+e^{-\gamma}E\right)^2}{(V^2+j^2\rho^4)^2},\\
\mathcal{P}&=-\frac{2 A_\varphi^\prime\left(4j A_\varphi+e^{-\gamma}E\right)}{(V^2+j^2\rho^4)\rho}.
\end{aligned}
\end{equation}
The Melvin-Bonnor-Swirling-ModMax spacetime configuration, here presented for the first time, reads 
\begin{widetext}
\begin{equation}\label{swirlingmelvinansatz}
\begin{aligned}
ds^2&=\frac{\rho^2}{V^2+j^2\rho^4}\left(d\varphi+4jz dt\right)^2+\left(V^2+j^2\rho^4\right)\left(-dt^2+d\rho^2+dz^2\right),\\
\mathcal{A}&=(e^{-\gamma}Ez+4jz A_\varphi)dt+A_\varphi d\varphi,
\end{aligned}
\end{equation}
where
\begin{equation}
\begin{aligned}
V&=1+\frac{e^{-\gamma} (E^2+B^2)}{4}\rho^2,\\
A_\varphi&=\frac{e^{-\gamma}B}{4j}\sin\left(e^\gamma\left[\pi-2\arctan\left(\frac{V}{j\rho^2}\right)\right]\right)+\frac{e^{-\gamma}E}{4j}\cos\left(e^\gamma\left[\pi-2\arctan\left(\frac{V}{j\rho^2}\right)\right]\right)-\frac{e^{-\gamma}E}{4j}.
\end{aligned}
\end{equation}
\end{widetext}
Recall that, in this case, the integrated function is the magnetic function $A_\varphi$, rather than $A_t$. This magnetic function induces an electric field of the desired form $\sim \omega A_{\varphi}$. Additionally, a background electric field of the form $e^{-\gamma}Ez$ serves the same role as the $q_m\cos\theta$ term in the Taub-NUT analog \eqref{ourTaub-NUT}.

Just as in the Taub-NUT case, the line element is only slightly modified by the inclusion of the screening factor 
$e^{-\gamma}$, which means a detailed analysis of the causal structure is not necessary, as it has been addressed in \cite{Barrientos:2024pkt}. However, the electromagnetic configuration does undergo significant modification. Nevertheless, the corresponding Einstein-Maxwell and Melvin-Bonnor-ModMax limits are properly recovered for 
$\gamma\rightarrow0$ and $j\rightarrow0$. In fact, the Einstein-Maxwell Melvin-Swirling spacetime of \cite{Barrientos:2024pkt,Capobianco:2024jhe} is recovered, as seen from the expansion
\begin{equation}
\mathcal{A}=\mathcal{A}_{(0)}+\gamma\mathcal{A}_{(1)}+\mathcal{O}(\gamma^2), 
\end{equation}
with
\begin{equation}
\begin{aligned}
\mathcal{A}_{(0)}&=\left[\frac{E(\bar{V}^2-j^2\rho^4)z}{\bar{V}^2+j^2\rho^4}+\frac{2jB\bar{V}\rho^2 z}{\bar{V}^2+j^2\rho^4}\right]dt\\
    &\quad +\left[\frac{B\bar{V}\rho^2}{2(\bar{V}^2+j^2\rho^4)}-\frac{jE\rho^4}{2(\bar{V}^2+j^2\rho^4)}\right]d\varphi,
\end{aligned}
\end{equation}
\begin{equation}
\begin{aligned}
\mathcal{A}_{(1)}&=-\mathcal{A}_{(0)}+\left(\frac{B(\bar{V}^2-j^2\rho^4)}{\bar{V}^2+j^2\rho^4}-\frac{2jE\bar{V}\rho^2}{\bar{V}^2+j^2\rho^4}\right)\\
&\quad\times \left(\frac{8j(E^2+B^2)\rho^4}{\bar{V}^2+j^2\rho^4}+\pi-2\arctan\left(\frac{\bar{V}}{j\rho^2}\right)\right)\\
&\quad\times\left( z dt+\frac{d\varphi}{4j}\right),
\end{aligned}
\end{equation}
and where $\bar{V}$ corresponds to the polynomial $V$ evaluated at $\gamma=0$. Furthermore, the Melvin-Bonnor-ModMax configuration constructed in Section \ref{secII} is easily retrieved in the $j=0$ expansion 
\begin{equation}
\mathcal{A}\sim e^{-\gamma}Ez dt+\frac{B\rho^2}{2V}d\varphi+\mathcal{O}(j), 
\end{equation}
the leading order corresponding to the desired gauge field producing the expression \eqref{melvinMODMAXgauge}. Indeed, the swirling geometry of pure vacuum is obtained via $(E,B)\rightarrow0$. 


\section{Conclusions and further prospects}\label{secVI}

In this paper, we have extended the study of black hole solutions within the framework of Einstein-ModMax theory. Building on the existing catalog of exact solutions and considering the prevalent limitations in the characterization of rotating solutions, we have constructed Melvin-Bonnor and vortex-like geometries. Specifically, we have derived a Melvin-Bonnor-ModMax background, along with its (A)dS extension, and used these insights to construct, for the first time, both Schwarzschild and C-metric Melvin-Bonnor black holes immersed in the ModMax theory. Although the properties of these spacetimes closely resemble their Einstein-Maxwell counterparts, and therefore a detailed re-examination of their causal structures is not required, we present two novel contributions. We identify a previously unnoticed aspect of the Schwarzschild-Melvin-Bonnor-ModMax solution \eqref{schMelvinMODMAX} and the C-metric solution \eqref{Cmetricmodmax}. In particular, we show that both solutions can be reformulated through two distinct novel Kerr-Schild-type representations. These representations not only offer a fresh perspective on these familiar spacetimes but also unveil structural features that were previously hidden in their traditional forms. By introducing these innovative Kerr-Schild frameworks, we provide new insights that may have significant implications for the study of charged black hole solutions. It is important to note that the significance of these results is partly due to the fact that these constructions remain equally valid in the standard electrovacuum case by simply taking the limit $\gamma \rightarrow 0$.

In the context of stationary solutions, we also present, for the first time, a vortex-like solution, referred to as the Melvin-Bonnor-Swirling background in ModMax. The necessary intuition for constructing this solution was drawn from the previously known Taub-NUT solutions in ModMax \cite{BallonBordo:2020jtw, Flores-Alfonso:2020nnd}, which for our purposes have been written in the convenient gauge \eqref{gaugenostro}.

There are several promising directions for further exploration of the Einstein-ModMax theory, building upon the work we have presented here. One interesting avenue is embedding static and accelerating black holes within the Melvin-Bonnor-Swirling-ModMax background. Based on our understanding, the challenges we have faced are primarily due to computational limitations rather than conceptual ones, leading us to believe that these black hole solutions belong to the spectrum of exact solutions in Einstein-ModMax theory.
Additionally, following the approach in \cite{Ayon-Beato:2024vph}, the solutions presented here can be generalized to include a scalar field with conformal coupling. Moreover, these solutions can be further extended by introducing a (super-)renormalizable potential that breaks the conformal symmetry of the scalar field action, as discussed in \cite{Ayon-Beato:2015ada}.

Another valuable direction would be the full generalization of the Ernst scheme \cite{Ernst:1967wx, Ernst:1967by} for ModMax theory. Although achieving a comprehensive formulation of Einstein-ModMax theory using complex Ernst potentials will likely require significant effort, investigating potential Lie point symmetries, such as those identified in more complex systems \cite{Dowker:1993bt, Ortaggio:2004kr}, may yield valuable insights.

Lastly, a major and highly nontrivial challenge lies in constructing rotating solutions within ModMax theory. We aim to report progress on this in future work.
\\
\acknowledgments
J.B. and A.C. appreciate interesting discussions with Crist\'obal Corral and Prof. Pavel Krtou{\v s}. We also thank Eloy Ay\'on-Beato for clarifying some important issues. The work of J.B. is supported by FONDECYT Postdoctorado grant 3230596. A.C. is partially supported by FONDECYT grant 1210500 and by PRIMUS/23/SCI/005 and GA{\v C}R 22-14791S grants from Charles University.  The work of M.H. is partially supported by FONDECYT grant 1210889. The authors would like to express their gratitude to the cities of Krakow, Trieste, Split, Athens, Sifnos and Milos, for providing the necessary environment to finish this work.

\bibliography{apssamp}

\end{document}